\documentclass[aps,prx,showpacs,twoside,twocolumn,10pt,longbibliography]{revtex4-2}
\usepackage[colorlinks=true, citecolor=red, urlcolor=blue ]{hyperref}
\usepackage{times,epsfig,amssymb,amsfonts,amsmath,bm,subfigure,mathtools,amsthm,braket,soul,enumitem,color, caption} %, subcaption
\usepackage[normalem]{ulem}
\usepackage{subcaption}
\usepackage{graphicx}
\usepackage{xcolor}
\usepackage{subfloat}
\usepackage{float}
\usepackage{braket}
\usepackage{tabularx}
\usepackage{lipsum}
\usepackage{booktabs}
\usepackage{multirow}
\newcolumntype{C}[1]{>{\centering\arraybackslash}p{#1}}

\newcommand{\stkout}[1]{\ifmmode\text{\sout{\ensuremath{#1}}}\else\sout{#1}\fi}
\definecolor{magenta}{rgb}{1.0, 0.0, 0.56}

\newcolumntype{M}[1]{>{\centering\arraybackslash}m{#1}}
\newcolumntype{N}{@{}m{0pt}@{}}

\begin{document}
\title{Permutation-equivariant quantum convolutional neural networks}
\author{Sreetama Das$^{1, 3}$, Filippo Caruso$^{1,2,3}$}
\affiliation{$^1$Istituto Nazionale di Ottica del Consiglio Nazionale delle Ricerche (CNR-INO), I-50019 Sesto Fiorentino, Italy}
\affiliation{$^2$Department of Physics and Astronomy, University of Florence, Via Sansone 1, Sesto Fiorentino, I-50019, Italy}
\affiliation{$^3$European Laboratory for Non-Linear Spectroscopy (LENS), University of Florence, Via Nello Carrara 1, Sesto Fiorentino, I-50019, Italy}
%\affiliation{$^3$Fakultaet Physik, Technische Universitaet Dortmund, D-44221 Dortmund, Germany}

\begin{abstract}
The Symmetric group $S_{n}$ manifests itself in large classes of quantum systems as the invariance of certain characteristics of a quantum state with respect to permuting the qubits. The subgroups of $S_{n}$ arise, among many other contexts, to describe label symmetry of classical images with respect to spatial transformations, e.g. reflection or rotation. Equipped with the formalism of geometric quantum machine learning, in this work we propose the architectures of equivariant quantum convolutional neural networks (EQCNNs) adherent to $S_{n}$ and its subgroups. We demonstrate that a careful choice of pixel-to-qubit embedding order can facilitate easy construction of EQCNNs for small subgroups of $S_{n}$. Our novel EQCNN architecture corresponding to the full permutation group $S_{n}$ is built by applying all possible QCNNs with equal probability, which can also be conceptualized as a dropout strategy in quantum neural networks. For subgroups of $S_{n}$, our numerical results using MNIST datasets show better classification accuracy than non-equivariant QCNNs. The $S_{n}$-equivariant QCNN architecture shows significantly improved training and test performance than non-equivariant QCNN for classification of connected and non-connected graphs. When trained with sufficiently large number of data, the $S_{n}$-equivariant QCNN shows better average performance compared to $S_{n}$-equivariant QNN . These results contribute towards building powerful quantum machine learning architectures in permutation-symmetric systems.
\end{abstract}

\maketitle

\section{Introduction}
 
The field at the interface of quantum technology and machine learning has been subject to extensive research in recent times. A significant fraction of these are aimed at building machine learning models using quantum systems. Of particular interest are quantum neural networks (QNNs)~\cite{qml_schuld_2015, qml_lloyd_2017, Benedetti_2019, Perdomo-Ortiz_2018, schuld_2022_prx}, analogous to classical neural networks in deep learning. The central component in a prototypical QNN is a quantum circuit with single and multiple qubit gates with trainable parameters. It has been widely known as parametric quantum circuit (PQC) or variational quantum circuit (VQC)~\cite{peruzzo_2014, McClean_2016, Romero_2017, mitarai_2018, schuld_2020}. PQC-based QNNs have been used to design quantum analogs of well-known classical machine learning architectures, e.g. quantum autoencoder~\cite{Romero_2017}, quantum convolutional neural networks (QCNNs)~\cite{cong_2019, Li_2020, quanvolution_2020}, quantum generative adversarial networks (QGANs)~\cite{Braccia_2021, braccia_2022, rudolph2022generation, boyle2023hybrid, tsang2023hybrid, ZHOU2023116891}, quantum generative diffusion models~\cite{parigi2023quantumnoisedriven,zhang2023generative,cacioppo2023quantum} etc. The cost function obtained from the above networks is optimized using a classical optimization routine, thus they are \textit{hybrid quantum-classical networks}. They are designed to function with both classical and quantum data. Despite the potential speed-up and signatures of success of quantum machine learning over classical ML models \cite{rebentrost_svm_2014, Liu_2021, abbas_2021, huang_science_2022, caro_2022}, there exist several limitations for QNNs, the \textit{barren plateau} problem being one of them~\cite{mcclean_barren_2018, holmes_2022, cerezo_2021, sharma_2021}. On one hand, an arbitrary \textit{hardware-efficient} PQC ansatze is expected to have high expressibility \cite{sim_2019} to ensure that the solution of the optimization problem is close enough to the actual solution. On the other hand, PQCs with higher expressibility are increasingly prone to exhibiting barren plateau~\cite{mcclean_barren_2018, holmes_2022}. It is therefore a crucial task to mitigate barren plateau for a practical application of quantum neural networks.

One way to improve the trainability and generalization in machine learning algorithms is to introduce \textit{inductive bias} in the learning model, i.e. to use some prior known information about the dataset to build a problem-specific model and constrain the optimization space of the network. Particularly, \textit{geometric machine learning} refers to a scheme in which the known symmetries of the dataset are used to construct a network which respects those symmetries~\cite{bronstein2021geometric}. In past few years, the theory of geometric quantum machine learning has been developed by a number of works~\cite{larocca_2022, mernyei2022equivariant, skolik2023equivariant, meyer_2023, nguyen_2022, zheng_2023, east2023need}. The performance of symmetry-enhanced QNNs, also called Equivariant QNNs (EQNNs), has been investigated for a number of commonly encountered symmetries in classical datasets as well as quantum systems. These studies show that some classes of EQNNs are devoid of the barren plateau problem~\cite{pesah_2021, schatzki2022theoretical, west2023provably}. Moreover, the performance of an EQNN shows improvement over a non-equivariant QNN for pattern recognition and image classification~\cite{meyer_2023, West_2023, chang2023approximately, west2023provably, das2024role}.

Among all the symmetry groups, the group of all permutations of $n$ objects, which is also called Symmetric group ($S_{n}$), is a crucial one. It is known that every group is isomorphic to some subgroup of a Symmetric group. Symmetric group frequently arises in physical scenarios where certain properties of a quantum state remain invariant under arbitrary permutation of the qubits. To exemplify, the genuine multiparty entangled quantum states remain genuinely entangled when the subsystems are permuted, or the set of connected graphs remain connected under permutation of the vertices. There also exist scenarios for which the permutation symmetry in a system corresponds to a subgroup of $S_{n}$. For example, the states of a many-body quantum system in presence of periodic boundary condition remain unchanged under cyclic permutation of qubits, which corresponds to cyclic group $Z_{n}$. Another example is the label symmetry of classical images, i.e. the class labels of images remain unchanged under spatial transformations like reflection or rotation, which corresponds to permuting the pixels in a certain way.
%, when each pixel value of the image has been encoded into a qubit (angle embedding). 
The $S_{n}$-equivariant QNN has been studied in \cite{schatzki2022theoretical}, with a theoretical proof of absence of barren plateau. For small subgroups of $S_{n}$ relevant to classical images, EQNN can be easily built by visualizing the qubits as pixels in a 2D plane~\cite{meyer_2023}.

While QNN broadly refers to a general PQC architecture, in this work, we focus our attention to a specific type of QNN, namely quantum convolutional neural networks. We propose an architecture of equivariant quantum convolutional neural network (EQCNN) for Symmetric group and its subgroups, which we jointly refer to as \textit{permutation group}.  Our architecture is based on the well-known model of QCNN introduced in Ref.~\cite{cong_2019}, which shows a considerably good performance for classifying topological phases of many-body quantum systems~\cite{cong_2019, herrmann_2022, liu_2023, umeano2023learn, monaco_2023} as well as classical images~\cite{Hur2022}. A few works have studied EQCNN for classical images~\cite{chang2023approximately, das2024role}, all of them using the quantum amplitude embedding. For reflection or rotation symmetry of images, amplitude embedding results in a representation of permutation group that acts locally on each qubit. However, when using qubit embedding (also known as angle embedding), the resulting representations correspond to subgroups of $S_{n}$ and act non-locally on multiple qubits, as we will elaborate in the next section. In addition to that, the QCNN proposed in Ref.~\cite{cong_2019} is subject to some structural restrictions, i.e. the translational invariance of the convolution layers and the qubit-reducing feature of pooling layers. The main challenge in designing EQCNN is to comply with these constraints under non-local action of the unitary representations in general. In this work, we try to resolve this issue. We demonstrate that for subgroups of $S_{n}$, it is important to choose a particular order of pixel-to-qubit mapping which enables one to build the EQCNN ansatze in a translationally-invariant manner. Equipped with this, we put forward EQCNN architectures for reflection symmetry and $\pi/2$-rotational symmetry of images, which shows improved performance compared to non-equivariant QCNNs for classifying MNIST datasets. For full permutation symmetry of $S_{n}$, our EQCNN is built by applying all QCNNs at random with equal probability. Such a probabilistic picture of QNN, although mentioned briefly in Ref.~\cite{nguyen_2022}, has not been explored significantly before. We build a quantum circuit to realize this stochastic action of the convolutional and pooling layers. We test its performance for classification of connected and non-connected graphs against non-equivariant QCNN as well as $S_{n}$-equivariant QNN. While the EQCNN always performs better than the former, for the latter it shows an improvement when trained with comparatively large number of data-points.

\section{Methods}
\label{prelim}
\subsection{Quantum Convolutional Neural Network (QCNN)}
The QCNN architecture of ref.~\cite{cong_2019} is inspired from the multi-scale entanglement renormalization ansatz (MERA) representation of quantum many-body states \cite{mera_2008}. Any QNN has three components: an $n$-qubit quantum register encoding the input quantum state $\vert \Psi\rangle$, a PQC $\mathcal{U}_\theta$ acting on $\vert \Psi\rangle$, and lastly a measurement $M$ on some (or all) of the qubits. In case of QCNN, $\mathcal{U}_\theta$ consists of a series of convolutional and pooling layers. The convolutional layer has a brick-like structure on an 1D array of qubits,  i.e. an $m$-qubit ($m<n$) trainable convolutional ansatz is applied on all combinations of neighboring $m$ qubits, mimicking the action of a kernel in CNN. It is a common practice to consider a two-qubit convolutional ansatze. All ansatze within the same convolutional layer have same trainable parameters. In the pooling layer, a subset of qubits are measured and conditioned on each measurement outcome, a parametrized rotation is applied on each qubit of the complementary subset. Similar to convolutional layers the trainable parameters are shared within a pooling layer. All the measured qubits are then traced out, thus reducing the effective dimension of the system. In real quantum devices this is same as ignoring the qubits in the subsequent stages and considering only the dynamics on the remaining qubits. The sequence of convolutional and pooling layers is then repeated until a small fraction of qubits are left. In analogy to the fully-connected part of a CNN, one can then apply a general PQC on the remaining qubits at the end and finally measure them to obtain the prediction. The network is then trained using a suitable loss function and optimization algorithm. Compared to a generic deep quantum neural network, in this case due to progressive qubits reduction, the structure has a shallower depth of $\mathcal{O}(\log n)$, which is preferable for implementation in NISQ devices. This model of QCNN is also known to be free of barren plateau~\cite{pesah_2021}.

\begin{figure*}[t]
    \centering
    %\hspace{2.8em}
    \includegraphics[width=0.75\textwidth]{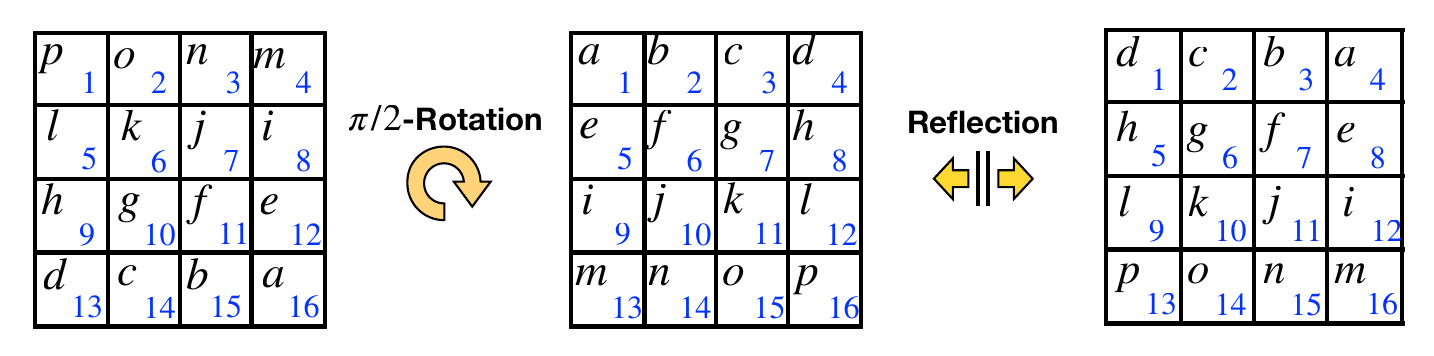}
    \caption{In the center, a $4\times 4$ image with pixel values $\{a, b, ..., p\}$ indicated in the top left corner of each pixel and corresponding pixel positions in the bottom right corner. On the right, the image after vertical reflection. On the left, the image after rotation by angle $\pi/2$. }
    %(b) The altered-order encoding of pixel values in qubits used for reflection equivariant EQCNN.}
    \label{ref_rot_schematic}
\end{figure*}

\subsection{Label symmetry}
For a classical or quantum dataset, a \textit{label symmetry} exists if the class labels of all the data-points remain unchanged under a set of operations on those data-points. For example, in MNIST dataset the labels of digits $\{0, 1, 8\}$ are unchanged when the images are reflected, or the labels of a quantum state being entangled or non-entangled are invariant under local unitary transformations on a qubit. Let us consider a binary-classification task for a dataset $\mathcal{X}$ with data-points $x_{i}$ and corresponding class labels $y_{i} \in \mathcal{Y} $ where $\mathcal{Y}=\{0, 1\}$. The function that maps the data-points to their labels is $f: \mathcal{X}\rightarrow \mathcal{Y}$. The task of QNN is to realize a quantum circuit $f_{\theta}$ that closely approximates $f$. A set of operations $\mathcal{G}$ form a label symmetry group of $\mathcal{X}$ if for all group elements $g\in\mathcal{G}$, the assigned labels remain unchanged, i.e.,
\begin{equation}
    f(g(x_{i})) = y_{i} = f(x_{i}), \quad \forall x_{i} \in \mathcal{X},\;\forall g\in \mathcal{G}.
\end{equation} 
Note that the data points themselves may not be invariant under the group action, i.e., $g(x_{i}) \neq x_{i}$ in general. Thus, the parametric function $f_{\theta}$ used to classify $\mathcal{X}$ will be consistent with $\mathcal{G}$ if it predicts same label for inputs related by the symmetry operations. The QNNs respecting this constraint are called Equivariant QNNs (EQNNs). It was proved in~\cite{nguyen_2022, ragone2023representation} that a PQC $\mathcal{U}_{\theta}$ and a measurement $M$ will form an EQNN if and only if
\begin{equation}
    [R(g), \mathcal{U}_\theta] = 0 \quad  \mbox{and} \quad [R(g), M] = 0.
    \label{eqv_condition2}
\end{equation}
Thus if we define the \textit{commutator space} of $R(g)$ to be the space of all operators that commute with $R(g)$, then $\mathcal{U}_{\theta}$ and $M$ must belong to that commutator space. In the context of EQCNN, $\mathcal{U}_{\theta}$ is composed of a series of convolutional and pooling layers. Thus, equivariance of $\mathcal{U}_{\theta}$ can be obtained by making sure that each convolution and pooling layer is equivariant with respect to $R(g)$.

\subsection{EQCNN with permutation symmetry}

The elements of a symmetry group can act either as \textit{inner symmetry} or \textit{outer symmetry}~\cite{nguyen_2022}. For the former, $R(g)$ can be decomposed as a tensor product of locally acting unitary operators. A well known example is the local unitary group $\mathrm{SU}(2)$ acting on a system of $n$ qubits, and its representation $R(g) = g^{\otimes n}$ which acts locally on each qubit, where $g \in \mathrm{SU}(2)$. Another commonly occurring instance of inner symmetry is when classical images are encoded as quantum states using amplitude embedding, and the representation $R(g)$ of reflection or rotation group symmetry is a tensor product of Pauli-$X$ matrices and qubit identity operators acting locally on different qubits. The architecture and performance of equivariant quantum neural networks for inner symmetries have been investigated in a number of works~\cite{meyer_2023, nguyen_2022, West_2023, chang2023approximately, das2024role}.

On the other hand, $R(g)$ for outer symmetry groups cannot be decomposed into locally acting components, an example being the Symmetric group $S_{n}$ which manifests itself as a symmetry with respect to permuting the qubits. The physical operation of permuting two qubits is expressed mathematically using SWAP gate. One can show that, for every element $g$ of $S_{n}$, $R_{g}$ can be written as a product of SWAP gates, as we will exemplify in the next section. When a system of $n$ qubits has full permutation symmetry, each qubit can be swapped with any other qubit. The fulfillment of equivariance condition Eq. (\ref{eqv_condition2}) then requires that $\mathcal{U}_{\theta}$ must remain unchanged when the qubits are permuted in an arbitrary way. When using layers of single-qubit gates and two-qubit entangling gates to construct $\mathcal{U}_{\theta}$, permutation equivariance can be achieved if one uses single qubit gates with same parameters on all qubits, and entangling gates to connect all possible ${n}\choose{2}$ combinations of two qubits. In reference Ref. \cite{schatzki2022theoretical}, the authors show that this architecture of $S_{n}$-equivariant QNN is devoid of barren plateau, and benchmark its performance for classification of connected and non-connected graph states.

This structure of $S_{n}$-equivariant QNN is translationally symmetric on an array of qubits, thus can be used as the first convolutional layer of a permutation-equivariant QCNN. However, it is not straightforward to construct pooling layers in this case, as choosing a particular set of qubits to trace out inevitably breaks the full permutation symmetry. Additionally, the reduced representation on the remaining qubits after the first pooling layer is not a faithful representation. It was pointed out in Ref. \cite{nguyen_2022} that one possible technique is to randomly trace out all combinations of $n\choose n/2$ qubits with equal probability to retain full permutation symmetry.
In this work, we have used this strategy to construct equivariant pooling layers.
%We demonstrate how to implement this strategy in a quantum circuit. 
Moreover, we show how to construct equivariant measurements for a probabilistic ansatze.

There exist other practically useful scenarios when the relevant labels of quantum states remain unchanged under the action of a subgroup of $S_{n}$. For example, when each pixel value of a classical image is encoded using a qubit (angle embedding), the operation of reflecting or rotating the image corresponds to permuting the qubits in a certain way, eliminating full freedom. In a way, this is an inner symmetry for which the group representation acts locally on a set of qubits which are allowed to permute within that set. In such cases, equivariant $\mathcal{U}_{\theta}$ can be achieved using less than ${n}\choose{2}$ two-qubit gates and a lower-depth circuit, as we will describe in the following subsections.

\begin{figure*}[ht]
    \centering
    %\hspace{2.8em}
    \subfigure[]{\includegraphics[width=0.5\textwidth]{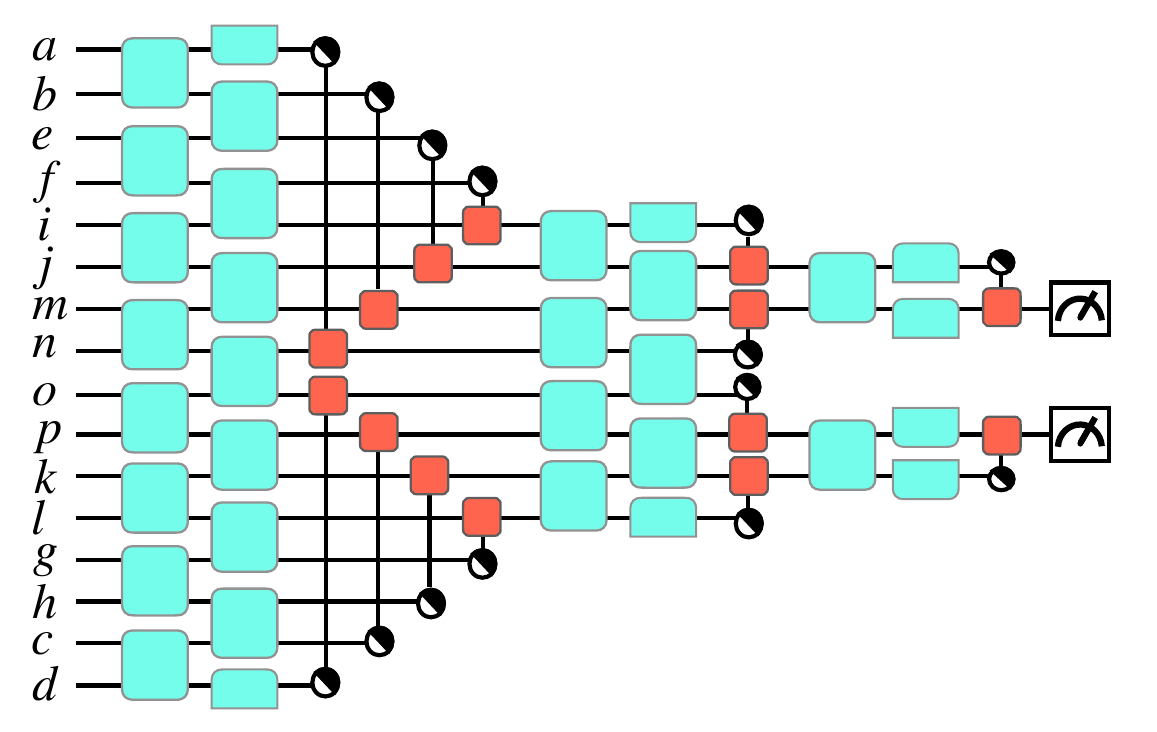}}\hspace{3em}
    \subfigure[]{\includegraphics[width=0.32\textwidth]{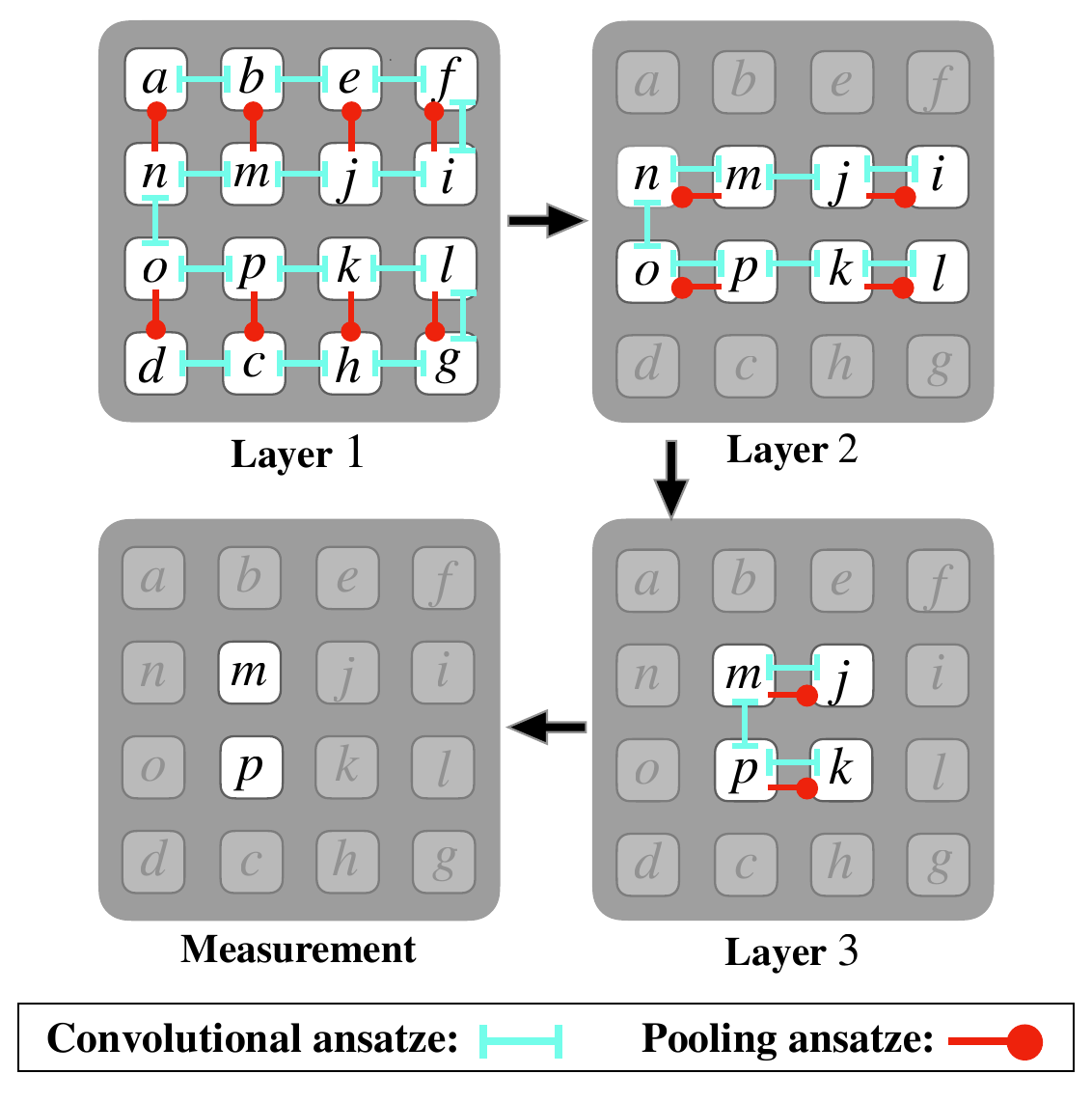}}
    \caption{(a) The structure of reflection-equivariant QCNN for 16 qubits. The cyan and the orange boxes represent respectively the parametrized convolutional and the pooling ansatze. For the control qubits in pooling layer, we use half-filled circles to indicate that the pooling ansatze depends on both the cotrol states $\{\vert 0\rangle, \vert 1\rangle\}$. (b) Reflection-equivariant QCNN using only nearest-neighbour connections in a 16-qubit quantum register with square-lattice architecture.}
    \label{eqcnn_refl}
\end{figure*}

\subsubsection{Reflection-equivariant QCNN}
\label{our_symmetry}
Let us consider classification of images whose labels remain invariant under a reflection about the vertical axis. There are two group elements here-- the identity operation $e$ which keeps the image unchanged, and the reflection operation $r$. They form the abstract group $\mathbb{Z}_{2} \equiv \{e, r\}$. 
For angle embedding of classical images into quantum states, the symmetry representation of reflection group acts non-locally on certain groups of qubits through SWAP gates. For example, for a $4\times 4$ image as in Fig. \ref{ref_rot_schematic}, the representation of the reflection group is,
\begin{widetext}
\begin{equation}
    R^{\mathrm{ref}} = \Big\{\mathbb{I}^{\otimes 16}, \; \mathrm{SWAP}_{1,4}\mathrm{SWAP}_{2,3}\mathrm{SWAP}_{5,8}\mathrm{SWAP}_{6,7}\mathrm{SWAP}_{9, 12}\mathrm{SWAP}_{10, 11}\mathrm{SWAP}_{13, 16}\mathrm{SWAP}_{14, 15}\Big\}
    \label{ref_repr}
    \end{equation}
\end{widetext}
where $\mathrm{SWAP}_{i,j}$ denotes swap operation between pixel values at positions $i$ and $j$ of the image before reflection. Here, qubit pairs $\{\{1, 4\}$, $\{2, 3\}..\}$ constitute subsystems on each of which a SWAP gate acts locally.
 
Now, for an arbitrary order of embedding the pixel values in 1D array of qubits, one can check that applying convolutional ansatze in the typical brick-like structure is not equivariant with respect to the representation in Eq. (\ref{ref_repr}). We demonstrate that by adhering to a particular pixel-to-qubit embedding order, one can build equivariant convolutional ansatze for all layers by having the brick-like arrangement of two-qubit ansatze. By pixel-to-qubit embedding order, we imply a particular one-to-one correspondence between pixel values $\{a, b, c, . ..\}$ and the qubits indices $\{1, 2, 3,..\}$ which encodes them. This pixel-to-qubit embedding order and the resulting EQCNN have been presented in Fig. \ref{eqcnn_refl} for a $4\times 4$ image. Here the qubit indices increase from 1 to 16 when moving from top to bottom of the 1D array. Due to this altered embedding order, the group representation changes to,
\begin{widetext}
\begin{equation}
    R^{\mathrm{ref}} = \Big\{\mathbb{I}^{\otimes 16}, \; \mathrm{SWAP}_{1,16}\mathrm{SWAP}_{2,15}\mathrm{SWAP}_{3,14}\mathrm{SWAP}_{4,13}\mathrm{SWAP}_{5, 12}\mathrm{SWAP}_{6, 11}\mathrm{SWAP}_{7, 10}\mathrm{SWAP}_{8, 9}\Big\}.
    \label{ref_repr_changed}
    \end{equation}
\end{widetext}
In the pooling layer, we trace out half of the qubits in a way so that the action of the reflection group has a faithful representation on the remaining qubits, which is the following after the first pooling layer,
\begin{equation}
    R^{\mathrm{ref}} = \Big\{\mathbb{I}^{\otimes 8},\; \mathrm{SWAP}_{5, 12}\mathrm{SWAP}_{6, 11}\mathrm{SWAP}_{7, 10}\mathrm{SWAP}_{8, 9}\Big\}.
\end{equation}
Similarly, after the second pooling layer, the reduced representation is
\begin{equation}
    R^{\mathrm{ref}} = \Big\{\mathbb{I}^{\otimes 4},\; \mathrm{SWAP}_{6, 11}\mathrm{SWAP}_{7, 10}\Big\},
\end{equation}
and that after the third pooling layer is
\begin{equation}
    R^{\mathrm{ref}} = \Big\{\mathbb{I}^{\otimes 2},\; \mathrm{SWAP}_{7, 10}\Big\}.
\end{equation}
The measurement is performed on the remaining two qubits $\{7, 10\}$.

This structure of EQCNN can be conveniently embedded on a quantum register with 2D square lattice architecture. We demonstrate this in Fig. \ref{eqcnn_refl}(b). An advantage of 2D is that all the pooling ansatze are applied on nearest-neighbour qubits. However, to retain full translation symmetry of the convolutional layer, one needs to apply the convolutional ansatze on two qubits $\{a, d\}$ which are not nearest neighbours in this case. Square lattice architectures have been realized in NISQ devices such as Rigetti quantum hardware ANKAA-2 and Pasqal neutral atom quantum processing units.

\begin{figure*}[ht]
    \centering
    %\hspace{2.8em}
    \includegraphics[width=0.70\textwidth]{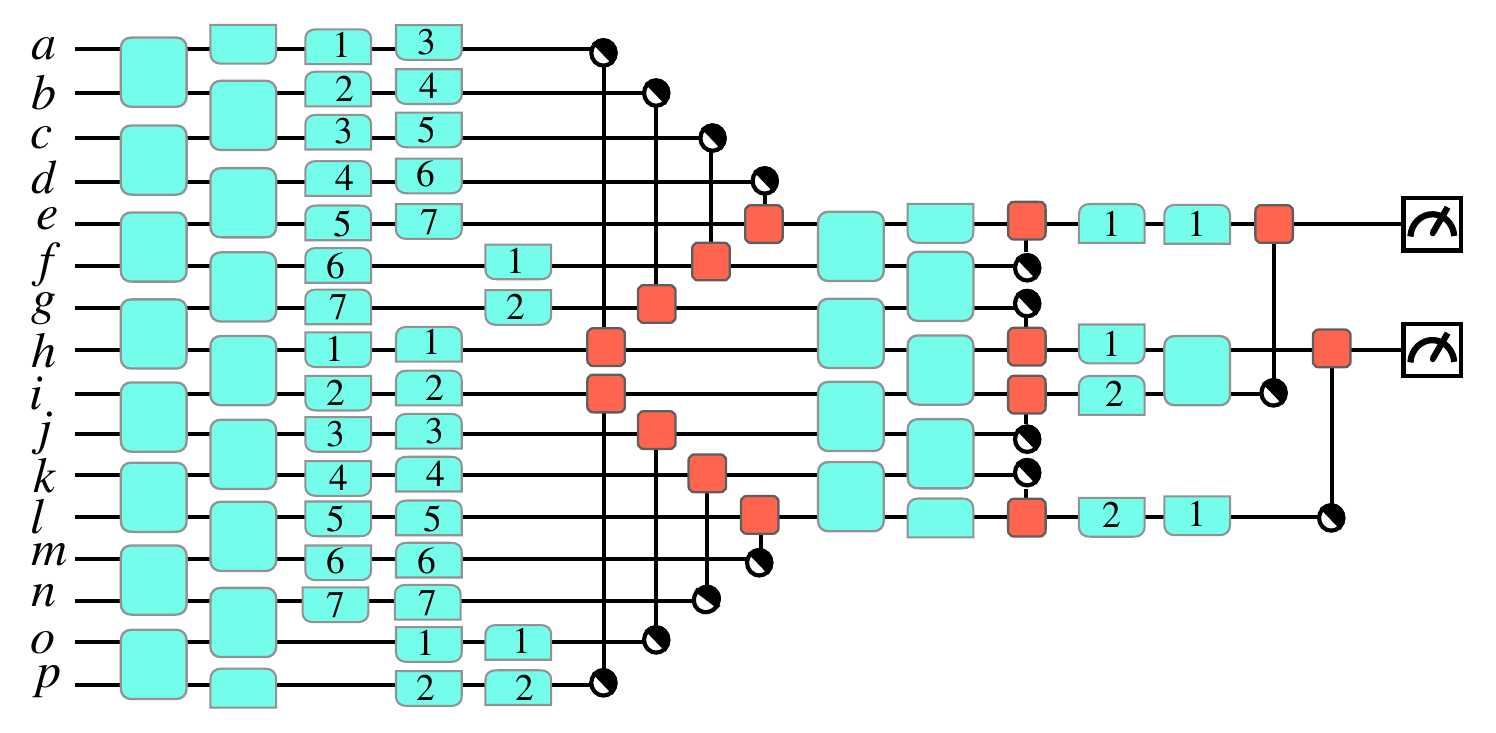}
    \caption{EQCNN for reflection and $\pi/2$-rotation symmetry for 16 qubits. For non-nearest-neighbour applications of two-qubit convolutional ansatze in the first and third layer, we use same integers to clearly indicate qubits on which they apply within a sub-layer .}
    \label{eqcnn_rot_ref}
\end{figure*}

\subsubsection{Reflection and $\frac{\pi}{2}$-rotation equivariant QCNN}
Next we consider images having label symmetry with respect to both reflection and a rotation by angle $\pi/2$.  We present the corresponding EQCNN in Fig. \ref{eqcnn_rot_ref}. There are two symmetry groups at play-- the reflection symmetry group $\{e, r\}$, and the rotational symmetry group $\{e, r^{\prime}\}$ where $r^{\prime}$ corresponds to $\pi/2$-rotation. For the natural order of pixel-to-qubit embedding, the group representation for the latter is
\begin{widetext}
\begin{equation}
    R^{\mathrm{rot}} = \Big\{\mathbb{I}^{\otimes 16}, \;\mathrm{SWAP}_{1,16}\mathrm{SWAP}_{2,15}\mathrm{SWAP}_{3,14}\mathrm{SWAP}_{4,13}\mathrm{SWAP}_{5, 12}\mathrm{SWAP}_{6, 11}\mathrm{SWAP}_{7, 10}\mathrm{SWAP}_{8, 9}\Big\},
    \label{ref_rot_repr}
    \end{equation}
\end{widetext}
as can be understood from Fig. \ref{ref_rot_schematic}. Note that Eq. (\ref{ref_rot_repr}) is the same as the representation in Eq. (\ref{ref_repr_changed}). Hence in this case, rotational equivariance is obtained from the brick-like structure of convolutional layer for the natural order of pixel-to-qubit embedding. However, to ensure reflection equivariance for this embedding, an additional layer of convolutional ansatze needs to be applied but in a different fashion. In detail, the two-qubit convolutional ansatz is applied on every qubit and and its $7^{\mathrm{th}}$ nearest neighbour when counting the qubits from top to bottom in Fig. \ref{eqcnn_rot_ref}. The reasoning behind this construction can be understood by observing how the nearest neighbour qubits are rearranged under reflection. The construction of first pooling layer ensures that the representations of both the groups on the remaining qubits are faithful. These reduced representations are
\begin{widetext}
    \begin{eqnarray}
        R^{\mathrm{ref}} = \Big\{ \mathbb{I}^{\otimes 8}, \; \mathrm{SWAP}_{5,8}\mathrm{SWAP}_{6,7}\mathrm{SWAP}_{9, 12}\mathrm{SWAP}_{10, 11}\Big\},\\
        R^{\mathrm{rot}}=\Big\{\mathbb{I}^{\otimes 8}, \; \mathrm{SWAP}_{5, 12}\mathrm{SWAP}_{6, 11}\mathrm{SWAP}_{7, 10}\mathrm{SWAP}_{8, 9}\Big\}.
    \end{eqnarray}
\end{widetext}
In the second convolutional layer, the brick-like structure alone is sufficient to ensure equivariance with respect to both the symmetries. The reduced representations after the second pooling layer are,
\begin{eqnarray}
    R^{\mathrm{ref}} = \Big\{\mathbb{I}^{\otimes 4},\;\mathrm{SWAP}_{5,8}\mathrm{SWAP}_{9, 12}\Big\},\\ R^{\mathrm{rot}}=\Big\{\mathbb{I}^{\otimes 4},\; \mathrm{SWAP}_{5, 12}\mathrm{SWAP}_{8, 9}\Big\}.
\end{eqnarray}
The third convolutional layer is constructed in the same way as the second convolutional layer. In the third pooling layer, it is impossible to trace out any two qubits, maintaining equivariance with respect to both the symmetry groups. Hence, we use a pooling layer for which the rotational equivariance is broken but the reflection equivariance is protected. Lastly, the remaining two qubits $\{5, 8\}$ are measured.

\begin{figure*}[t]
    \centering
    \subfigure[]{\includegraphics[width=0.52\textwidth]{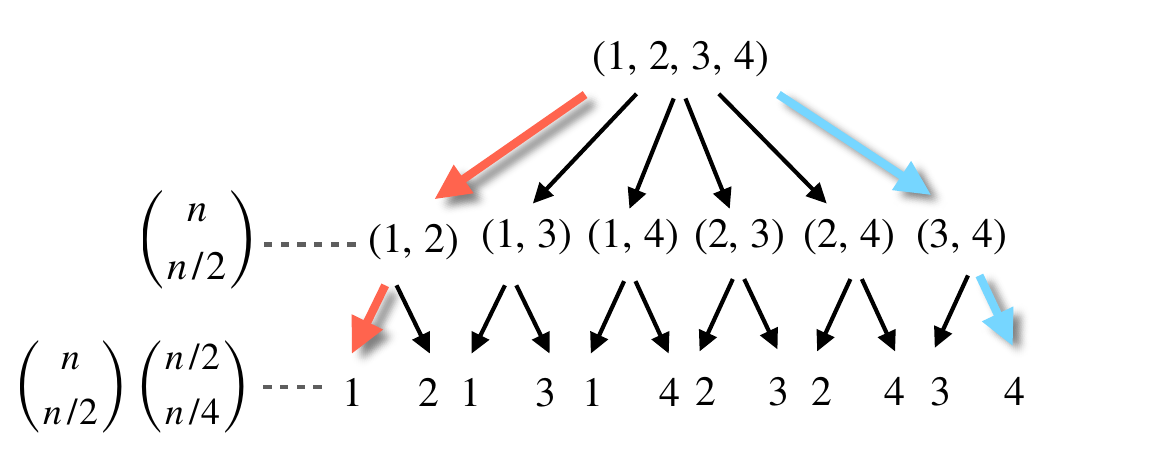}}\hspace{3em}
    \subfigure[]{\includegraphics[width=0.28\textwidth]{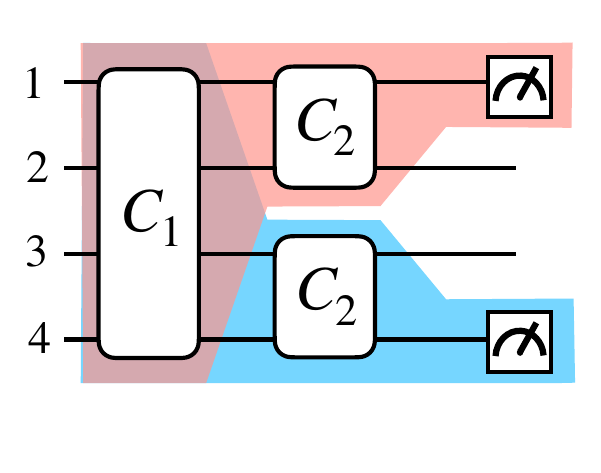}}
    \caption{(a) A schematic diagram showing all possible QCNNs and the qubit indices on which they act for an $S_{4}$-equivariant QCNN. Each arrow indicates a possible pooling operation. The number of different pooling operations in a layer is shown in the vertical column on the left. We highlight two examples of these QCNNs with red and blue arrows. (b) A circuit representation of the red and blue QCNNs and corresponding measurements. $C_{1}$ and $C_{2}$ are respectively the first and second convolutional layer.}
    \label{sn_conv_example}. 
\end{figure*}

Note that, one can obtain the EQCNN with respect to $\pi/2$-rotation symmetry alone by removing the additional convolutional layer in the first layer, and changing the last pooling ansatze in a way to protect rotational equivariance. As a result, the final measurement will be on a different set of qubits.

\begin{figure*}
    \centering
    %\hspace{2.8em}
    \includegraphics[width=0.80\textwidth]{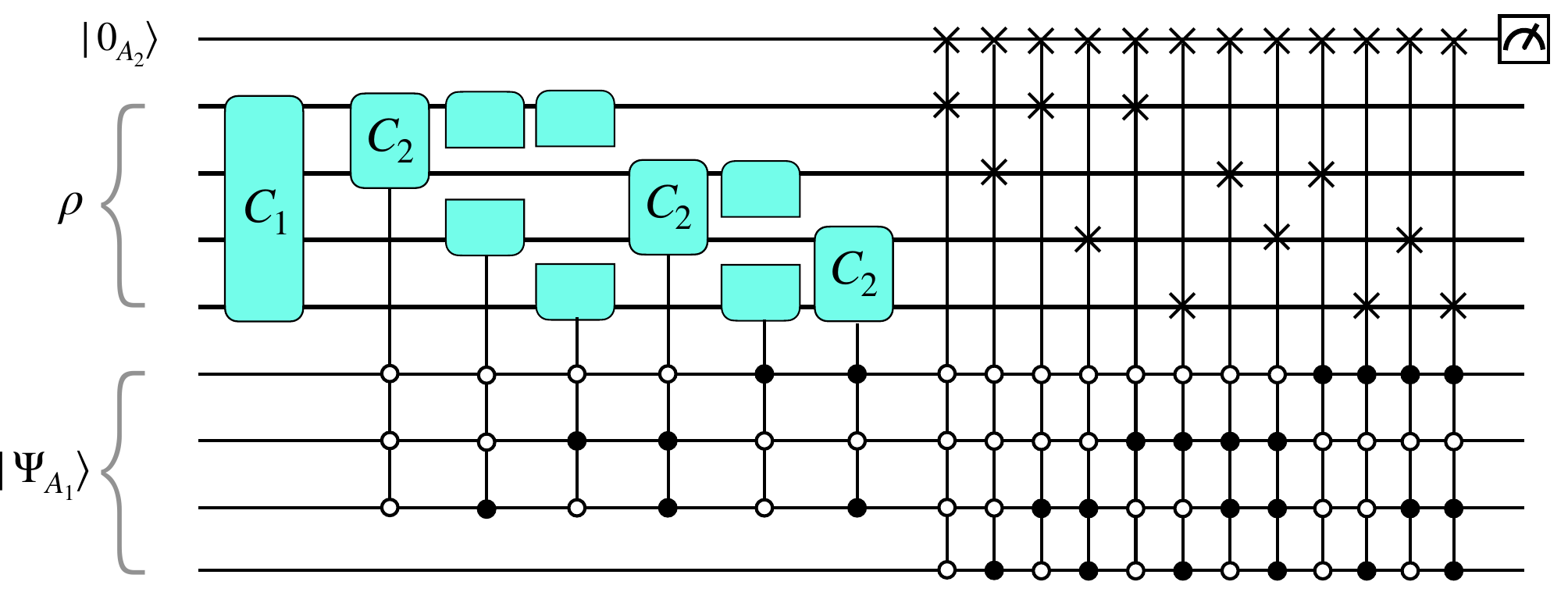}
    \caption{The circuit for realizing EQCNN for symmetric group $S_{4}$ on the input state $\rho$. In the controlled gates, the circles with no fillings and black fillings represent states $\vert 0\rangle$ and $\vert 1\rangle$ respectively.}
    \label{eqcnn_s4}
\end{figure*}

\begin{figure}
    \centering
    %\hspace{2.8em}
    \includegraphics[width=0.48\textwidth]{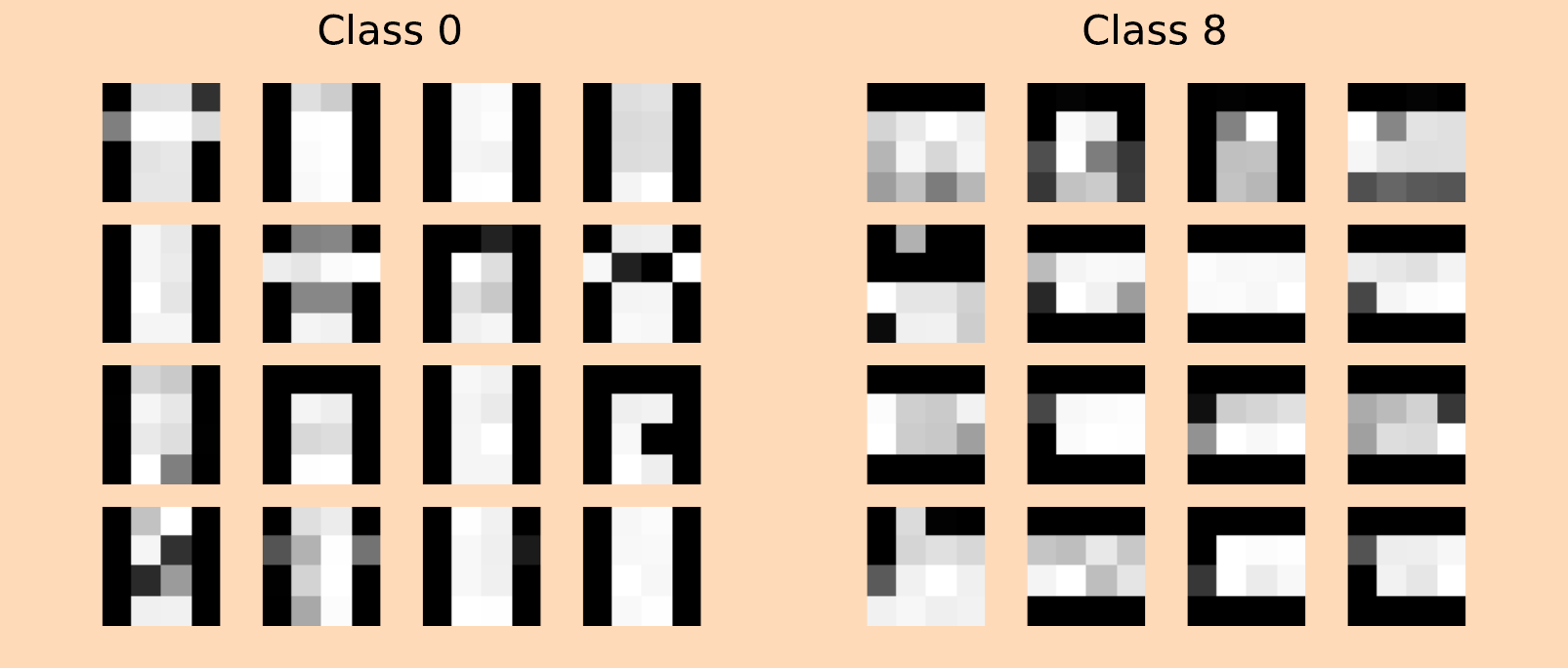}
    \caption{Samples of Fashion MNIST images corresponding to class 0 and class 8 sized down to $4\times 4$ pixels.}
    \label{fmnist_data}
\end{figure}

\begin{figure*}[!htb]
    \centering       
    \includegraphics[width=0.75\linewidth]{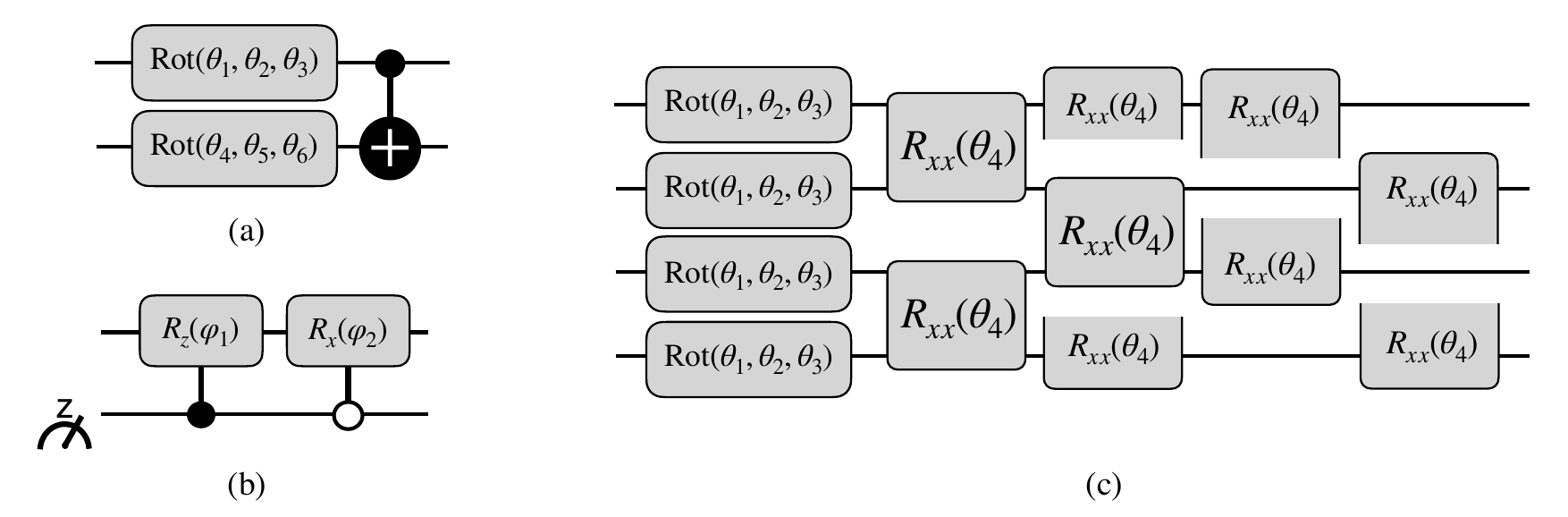}     
    \caption{(a) The convolutional ansatze and (b) the pooling ansatze used to construct EQCNN and non-equivariant QCNN for reflection and $\pi/2$-rotation symmetry. (c) The convolutional ansatze used in $S_{n}$-equivariant QCNN.}
    \label{ansatze}
\end{figure*}

\subsubsection{$S_{n}$-equivariant QCNN}
When a system has full permutation symmetry, we use the $S_{n}$-equivariant QNN with all-to-all two-qubit gate connections as the first convolutional layer. For the pooling operation, let us consider a scenario in which no explicit pooling ansatz is applied, but half of the qubits are traced out successively at each pooling layer. In the first pooling layer, we randomly trace out any $n/2$ qubits. This can be done in $n\choose n/2$ possible ways, each occurring with equal probability to retain the permutation equivariance. The second convolutional layer has the same structure as the first convolutional layer, however it is applied on the remaining $n/2$ qubits.
In the second pooling layer, we randomly trace out $n/4$ qubits from $n/2$ qubits in $n/2 \choose n/4$ possible ways, and apply the third convolutional layer on the remaining $n/4$ qubits. We keep adding convolutional and pooling layers until only one qubit is left. Another way to envision this is to think that there are in total
\begin{equation}
    P = \binom{n}{n/2} \times \binom{n/2}{n/4} \times \binom{n/4}{n/8}\times...\times \binom{2}{1}
    \label{number_eqcnn}
\end{equation}
 number of possible QCNNs, which are applied on the input state with equal probability. 
 %Note that $P$ is exponential in $n$. 
 For each of these QCNNs, the last remaining qubit is measured to obtain the loss function. We schematically depict the idea discussed above in Fig. \ref{sn_conv_example} for a simple case of $n=4$.

Now we build a quantum circuit that can exactly realize the probabilistic nature of the EQCNN described above. We show this in Fig. \ref{eqcnn_s4}. Additional to the target register encoding the input state $\rho$, there is an auxiliary quantum register $A_{1}$ with $\lceil\log P\rceil = 4$ qubits, which is initialized in an equal superposition $\vert \Psi_{A_{1}}\rangle$ of the first $P = 12$ orthogonal basis states. We apply the first convolutional layer $C_{1}$ on $\rho$.
Since no trainable pooling ansatz is present, tracing out $n/2$ randomly chosen qubits in the first pooling layer is equivalent to applying the second convolutional layer $C_{2}$ on all possible $\binom{4}{2} = 6$ pairs of qubits. To ensure that they are applied at random with equal probability, we apply them controlled on the states of the first three qubits in $A_{1}$. In the next and final pooling layer, randomly tracing out any one qubit from a qubit pair is equivalent to randomly choosing a qubit and measuring it. For this, we add another single-qubit auxiliary register $A_{2}$ initialized in state $\vert 0\rangle$. Now for each qubit pair in the target register, controlled on the last qubit of $A_{1}$ being in state $\vert 0\rangle$ or $\vert 1\rangle$, we swap either of the qubits with $A_{2}$. Finally, we measure Pauli-Z operator on $A_{2}$ to obtain the loss function. A detailed mathematical description of how this circuit works is discussed in the Appendix. 

A drawback of the above circuit is that with increasing $n$, $P$ increases very fast and soon becomes exponential in $n$, resulting in an exponentially deep circuit. This can be avoided if instead of tracing out half of the qubits from previous layer, one chooses to trace out a constant number $m$ of qubits at each pooling layer, and repeats this for $k <\frac{n}{m}$ times such that $P$ is polynomial in $n$. At the end of $k^{\mathrm{th}}$ layer, measurements are performed on all the remaining qubits. This strategy remains valid since tracing out exactly half of the qubits is a convention rather than a strict requirement for QCNN, although the logarithmic depth may not hold depending on the relation between $m$ and $n$. We leave it for future research works to investigate the performance of this polynomially deep EQCNN.

As mentioned in ref.~\cite{nguyen_2022}, the technique of probabilistically applying different unitary ansatze on the input states can also be used to design a dropout mechanism in a general quantum neural network. A few works exist in literature that proposes quantum dropout by randomly dropping single and two-qubit gates~\cite{kobayashi_2022, scala_2023} as well as the qubits themselves~\cite{schuld_2020}, whereas our strategy matches well with that discussed in ref.~\cite{verdon2018universal}. Particularly, ours is a special case in which half of the qubits from the previous layer are randomly chosen and dropped along with all gates acting on them on subsequent layers.  However, it also implies dropping half of the trainable parameters at each layer, which greatly reduces the expressibility and can result in poor trainability of the network. Instead, one may choose to drop a small constant number of $m$ qubits at each successive layer, as discussed also in the previous paragraph.

\begin{figure*}
    \centering
    \subfigure[]{\includegraphics[width=0.40\textwidth]{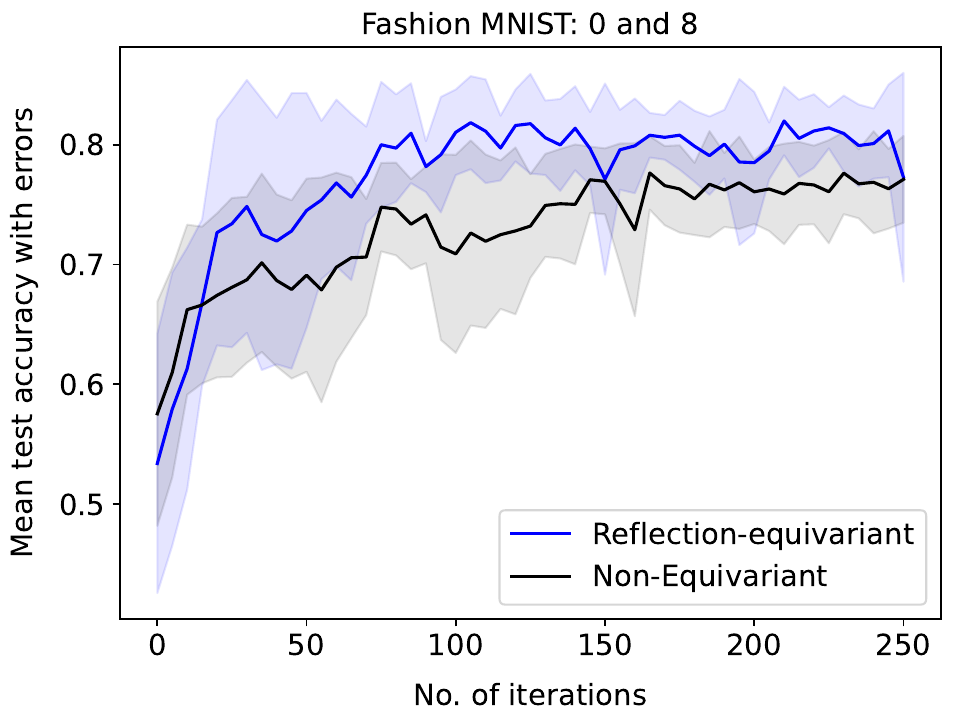}}\hspace{2em}
    \subfigure[]{\includegraphics[width=0.40\textwidth]{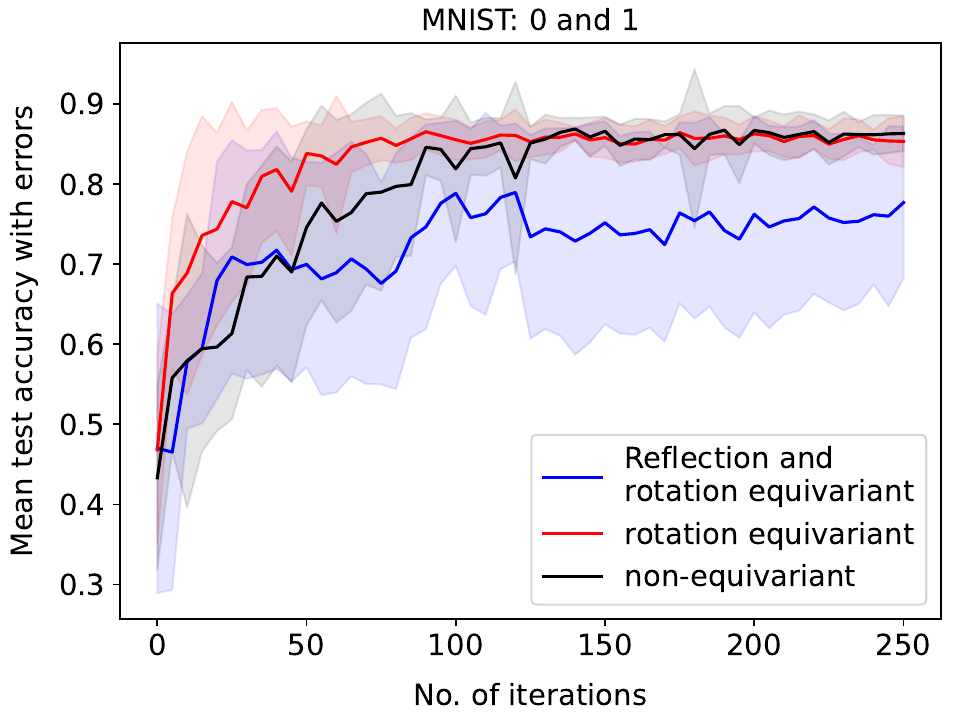}}
    \caption{Comparison of EQCNN and non-equivariant QCNN for binary classification of (a) classes $0$ and $8$ of Fashion MNIST dataset and (b) classes $0$ and $1$ of MNIST dataset. All quantities are averaged over 10 training iterations with random initial parameters.}
    \label{compare_accuracy}
\end{figure*}

\begin{figure*}
    \centering
    \subfigure[]{\includegraphics[width=1\textwidth]{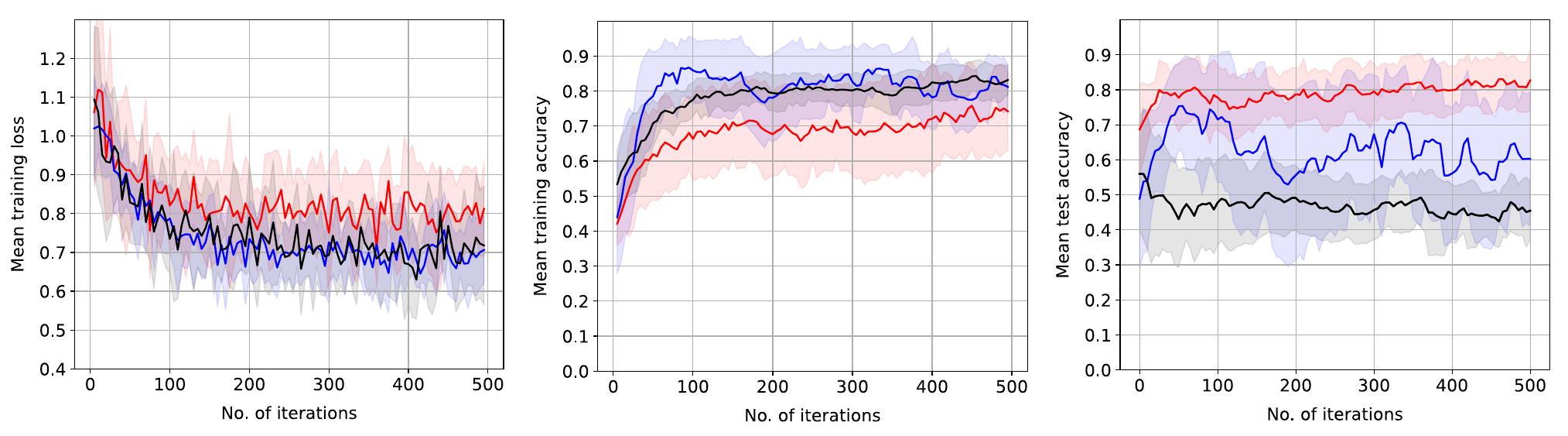}}
    \subfigure[]{\includegraphics[width=1\textwidth]{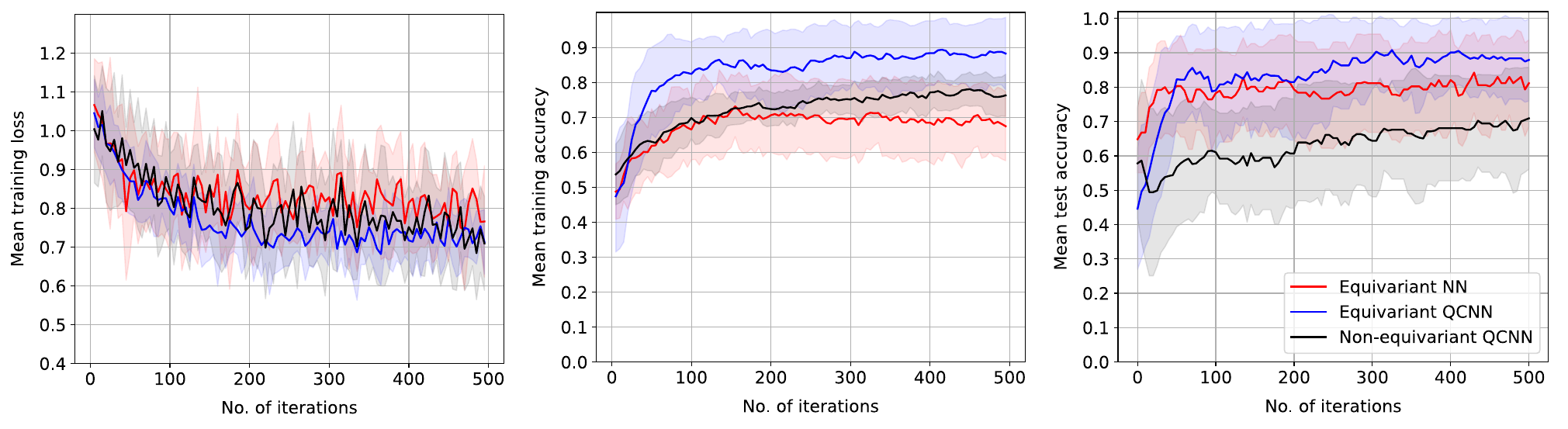}}
    \caption{Comparison of training set loss and accuracy, and test set classification accuracy of EQCNN, EQNN, and non-equivariant QCNN for classification of connected and non-connected graphs with $4$ vertices. All quantities are averaged over 10 training iterations with random initial parameters. (a) The training set and test set has respectively 45 and 18 distinct data-points. (b) The training set and test set has respectively 52 and 11 distinct data-points.}
    \label{compare_accuracy_sn}
\end{figure*}

\begin{figure}
    \centering
    %\hspace{2.8em}
    \includegraphics[width=0.45\textwidth]{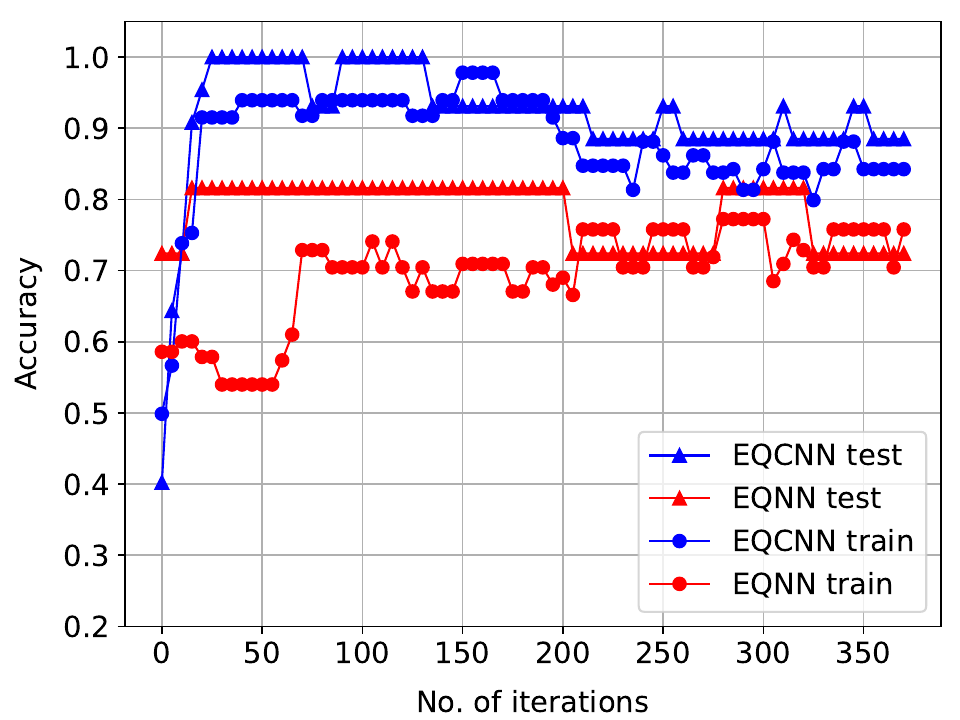}
    \caption{An instance of initial parameters for which $S_{n}$-EQCNN performs better than $S_{n}$-EQNN when the train-test split has $5:1$ ratio.}
    \label{two_plots_comparison}
\end{figure}

\section{Results}

We compare the classification accuracy of equivariant and non-equivariant QCNNs for the symmetry groups discussed in the last section. We build and train all the networks using Pennylane quantum simulator~\cite{bergholm2022pennylane}.

For reflection symmetry alone, we use class 0 (tshirt) and class 8 (handbag) of Fashion MNIST dataset, whereas for images symmetric with respect to reflection and $\pi/2$-rotation, we use class 0 and 1 of MNIST dataset. The original images are $28 \times 28$ pixels, we downsize them to $4\times 4$ pixels to embed the pixel values into 16 qubits using $R_{Y}$ rotation. Some samples of the downsized images for Fashion MNIST dataset are presented in Fig. \ref{fmnist_data}. As the convolutional and pooling ansatze, we use the circuits in Fig. \ref{ansatze}(a) and \ref{ansatze}(b) respectively, where $\mathrm{Rot}(\theta_{1}, \theta_{2}, \theta_{3}) = R_{Z}(\theta_{3})R_{Y}(\theta_{2})R_{Z}(\theta_{1})$. One can choose any other set of single-qubit and two-qubit gates without loss of generality. For building EQCNNs we follow the embeddings and architectures in Fig. \ref{eqcnn_refl} and Fig. \ref{eqcnn_rot_ref}, while for non-equivariant QCNNs we use standard pixel-to-qubit embedding order and the architectures in Ref. \cite{Hur2022}. We train the models using Nesterov moment optimizer with learning rate 0.005. As loss function, we use binary cross-entropy. We obtain the average test-set classification accuracy over $10$ identical runs, each with random initial parameters and a batch size of $32$. The behaviour of mean test-set accuracy with increasing number of training iterations is presented in Fig. \ref{compare_accuracy}. For fashion MNIST, EQCNN has a higher classification accuracy than non-equivariant QCNN. For MNIST images, considering the $\pi/2$-rotational symmetry alone results in a higher accuracy than the non-equivariant QCNN. However, taking into account both reflection and $\pi/2$-rotational symmetry reduces the accuracy significantly. This can be due to the highly constrained structure of EQCNN under both the symmetries, for which the expressibility degrades significantly.

We apply the $S_{n}$-equivariant QCNN for classification of connected and non-connected graphs with $4$ vertices. The graphs are randomly generated using Erd\H{o}s-R\'{e}nyi model with an edge probability of $45\%$. The quantum graph states are generated from the graphs using the standard method from~\cite{rauseendorf_2003, hein_2004}, and are embedded into a four-qubit quantum register using amplitude embedding. The property of being a connected or non-connected graph remains unchanged under arbitrary permutations of the vertices, in this case the qubits. From all possible graph states with $4$ vertices, we choose a fraction for training and the other part for testing. We use two different instances of this partitioning- for \emph{case 1} we have 45 data-points for training and 18 for testing, while for \textit{case 2} we have 52 and 11 data-points respectively for training and testing. In both cases, the training data is well balanced between the two classes, while the test data has a large fraction of connected graphs. We augment the training and test set with multiple copies of the distinct data-points.
As the convolutional ansatze, we use the $S_{n}$-equivariant ansatze shown in Fig. \ref{ansatze}(c) in the first layer, and its two-qubit analog in the second layer. The choice of the gates is again arbitrary. In addition to EQCNN, we also test the performance of $S_{n}$-equivariant QNN without pooling layer, for which we replace the second convolutional layer by another layer of the four-qubit ansatze in Fig. \ref{ansatze}(c). In this case we measure Pauli-Z operator on all the qubits to obtain the loss function. Note that the number of trainable parameters is same for the EQCNN and EQNN. For non-equivariant QCNN, we use the architecture of QCNN in Ref. \cite{Hur2022} with Fig. \ref{ansatze}(a) and \ref{ansatze}(b) as convolutional and pooling ansatze. All models are trained using Adam optimizer with a learning rate 0.01, and we use mean-squared error as loss function. We obtain the mean training loss, training accuracy and test accuracy from 10 randomly initialized training schemes with a batch size $10$. The results are presented in Fig. \ref{compare_accuracy_sn}.

On average, we find that the $S_{n}$-EQNN has a good generalization behaviour, and the corresponding test accuracy is better than the other two models for case 1 when the train-test ratio is $5:3$ approximately. The performance of $S_{n}$-EQNN does not differ much for case 1 and case 2. On the other hand, the $S_{n}$-EQCNN overfits the training data in case 1. However, when the train-test ratio is approximately $5:1$ in case 2 and there are more data-points to learn, the performance of the EQCNN improves and it shows higher average test accuracy than $S_{n}$-equivariant QNN. In this case, out of the 10 instances of initial parameters, there are more cases for which the maximum test accuracy $1$ is obtained for EQCNN compared to EQNN. In Fig. \ref{two_plots_comparison} we show the training and test accuracy for one such instance. The non-equivariant QCNN has a much lower test accuracy than both the equivariant architectures in both cases.

\section{Discussions and Conclusion}

In this work, we proposed a pixel-to-qubit embedding order that facilitates applying the quantum convolutional ansatze in an equivariant manner for angle-embedded classical images with reflection and rotational symmetry of image labels. The resulting EQCNNs show better classification accuracy than non-equivariant QCNNs. Though angle embedding requires as many qubits as the number of pixels compared to the logarithmically less number of qubits in amplitude embedding, the latter needs an exponentially deep circuit to encode an arbitrary quantum superposition of $n$ qubits, thus subsuming the advantage due to logarithmically shallow circuit of QCNN. Compared to that, angle embedding requires a circuit of constant depth, since each qubit can be manipulated individually in parallel. With this advantage, and the fast growing dimension of NISQ devices, angle embedding remains a more potent choice for quantum embedding of classical data. Moreover, one can also pre-process a large classical image to reduce its dimension using principal component analysis (PCA) or a classical autoencoder, before angle-embedding it into a smaller number of available qubits. In this case however, the pre-processing must preserve the label symmetry, in order to use an EQCNN for classification. It is to be noted that, though our embedding trick works for small subgroups of $S_{n}$, it will become increasingly difficult to design a translationally symmetric equivariant convolutional ansatze for bigger permutation groups, e.g. $p4m$ symmetry group, which again arises frequently in highly symmetric classical images. In such cases, one may choose to take into account fewer symmetries. This is supported by our numerical data showing a better performance compared to equivariant circuit which takes into account all existing symmetries. One can also use a more flexible ansatze that weakly breaks the symmetries~\cite{chang2023approximately}.

Our construction of $S_{n}$-equivariant QCNN proposed also a probabilistic picture of QNN, which has not been explored in detail in literature yet. We use graph states as dataset to analyze the performance of the $S_{n}$-equivariant EQCNN, which shows a better performance than non-equivariant QCNN. When sufficient training data are used, this EQCNN performs better than $S_{n}$-equivariant QNN on average. This probabilistic application of $S_{n}$-equivariant QCNN can be viewed as a special type of dropout mechanism in QNN~\cite{verdon2018universal, schuld_2020, kobayashi_2022, scala_2023}, in analogy to dropout in CNN to prevent overfitting~\cite{JMLR:v15:srivastava14a}. Particularly in our case, the important difference with the conventional dropout is that, here no trainable parameters are dropped, rather some of the qubits on which the corresponding gates apply. One can also apply with equal probability the equivariant ansatze for every symmetry group when there exist more than one symmetry group, as also discussed in the last paragraph.

It should be mentioned that there exist alternative approaches for designing a quantum convolutional neural network in quantum systems~\cite{quanvolution_2020, zheng_2023, zheng2022superexponential}. Particularly in references \cite{zheng_2023, zheng2022superexponential}, the authors conceptualize the qubit permutation symmetry as the quantum analog of translational symmetry in classical images. They propose as `convolutional quantum ansatze' a permutation-equivariant ansatze which is in turn constructed by employing global $SU(d)$-equivariant Hamiltonians, since by Schur's lemma the representation theory of $S_{n}$ can be described using the representation theory of $SU(d)$, and vice versa. As a result, their architecture can be used to find the ground states of many-body Hamiltonians that exhibit $SU(d)$ symmetry. This is in contrast to our work which proposes a permutation-equivariant QCNN that manifests the translation symmetry in a different way~\cite{cong_2019}.

Data classification is an ubiquitous task in a plethora of everyday applications. Classical state-of-the-art classifiers are supremely successful, yet the potency of quantum classifiers are worth the investigation. The quantum convolutional neural networks are proved to be efficient classifiers. Thus, our construction of equivariant QCNNs for spatial symmetry of images is a significant addition to the growing literature. The $S_{n}$-equivariant QCNN proposed in this work is extremely relevant due to the frequent presence of permutation symmetry in nature. We also believe that the stochastic nature of the variational quantum ansatz is an interesting direction to be pursued for future research.

\acknowledgments
This work was supported by the European Commission’s Horizon Europe Framework Programme under the Research and Innovation Action GA n. 101070546–MUQUABIS, by the European Union’s Horizon 2020 research and innovation programme under FET-OPEN GA n. 828946–PATHOS, by the European Defence Agency under the project Q-LAMPS Contract No B PRJ- RT-989, and by the MUR Progetti di Ricerca di Rilevante Interesse Nazionale (PRIN) Bando 2022 - project n. 20227HSE83 – ThAI-MIA funded by the European Union - Next Generation EU.\\

\appendix

\section{$S_{n}$-equivariant channel and measurements}
Denoting the input state as $\rho$, the state after the first convolutional layer is
\begin{equation}
    \rho_{1} = C_{1}\rho C_{1}^{\dagger},
\end{equation}
where $C_{1}$ is the unitary corresponding to the first convolutional layer.

The initial state in the auxiliary register $A_{1}$ is
\begin{eqnarray}
    &\vert \Psi_{A_{1}}\rangle = \frac{1}{\sqrt{12}}\big( \vert 0000\rangle + \vert 0001\rangle + \vert 0010\rangle + \vert 0011\rangle +\vert 0100\rangle \nonumber \\
    & + \vert 0101\rangle + \vert 0110\rangle + \vert 0111\rangle + \vert 1000\rangle + \vert 1001\rangle \nonumber \\
    &+\vert 1010\rangle + \vert 1011\rangle \big).
\end{eqnarray}
After the controlled application of second convolutional layer on different pairs of qubits, the input state becomes,
\begin{eqnarray}
    &\rho_{2} = \frac{1}{6} \big( C_{2}^{12} \rho_{1} C_{2}^{12^{\dagger}} + C_{2}^{13} \rho_{1} C_{2}^{13^{\dagger}} + C_{2}^{14} \rho_{1} C_{2}^{14^{\dagger}} \nonumber \\
    &+ C_{2}^{23} \rho_{1} C_{2}^{23^{\dagger}} + C_{2}^{24} \rho_{1} C_{2}^{24^{\dagger}} + C_{2}^{34} \rho_{1} C_{2}^{34^{\dagger}}\big) ,
\end{eqnarray}
where $C_{2}^{ij}$ is the second convolutional layer acting on the qubit-pair $(i, j)$.

At this stage, the auxiliary register $A_{2}$ is in the state $\vert 0\rangle$, and we denote the corresponding qubit with index $0$. After application of controlled swap gates, the state of the joint system consisting of our input state register and auxiliary register $A_{2}$ is,
\begin{widetext}
\begin{equation}
\begin{split}
    &\rho_{3}^{\prime} = \frac{1}{12}\big( S_{01}(C_{2}^{12} \rho_{1} C_{2}^{12^{\dagger}} \otimes \vert 0\rangle\langle 0|)S_{01}^{\dagger} 
    + S_{02}(C_{2}^{12} \rho_{1} C_{2}^{12^{\dagger}} \otimes \vert 0\rangle\langle 0|)S_{02}^{\dagger} + S_{01}(C_{2}^{13} \rho_{1} C_{2}^{13^{\dagger}} \otimes \vert 0\rangle\langle 0|)S_{01}^{\dagger}\nonumber \\
    & + S_{03}(C_{2}^{13} \rho_{1} C_{2}^{13^{\dagger}} \otimes \vert 0\rangle\langle 0|)S_{03}^{\dagger} + S_{01}(C_{2}^{14} \rho_{1} C_{2}^{14^{\dagger}} \otimes \vert 0\rangle\langle 0|)S_{01}^{\dagger} + S_{04}(C_{2}^{14} \rho_{1} C_{2}^{14^{\dagger}} \otimes \vert 0\rangle\langle 0|)S_{04}^{\dagger} \nonumber \\
    &+ S_{02}(C_{2}^{23} \rho_{1} C_{2}^{23^{\dagger}} \otimes \vert 0\rangle\langle 0|)S_{02}^{\dagger} + S_{03}(C_{2}^{23} \rho_{1} C_{2}^{23^{\dagger}} \otimes \vert 0\rangle\langle 0|)S_{03}^{\dagger} + S_{02}(C_{2}^{24} \rho_{1} C_{2}^{24^{\dagger}} \otimes \vert 0\rangle\langle 0|)S_{02}^{\dagger}  \nonumber \\
    &+ S_{04}(C_{2}^{24} \rho_{1} C_{2}^{24^{\dagger}} \otimes \vert 0\rangle\langle 0|)S_{04}^{\dagger} + S_{03}(C_{2}^{34} \rho_{1} C_{2}^{34^{\dagger}} \otimes \vert 0\rangle\langle 0|)S_{03}^{\dagger} +S_{04}(C_{2}^{34} \rho_{1} C_{2}^{34^{\dagger}} \otimes \vert 0\rangle\langle 0|)S_{04}^{\dagger}\big).
    \label{append_eq_3}
\end{split}
\end{equation}
\end{widetext}
The expectation value of the measurement on $A_{2}$ is,
\begin{eqnarray}
    m = \mathrm{Tr}(\rho_{3}^{\prime}M_{0}^{\prime}),
\end{eqnarray}
where $M_{0}^{\prime} = \sigma^{z}_{0}\otimes \mathbb{I}_{1} \otimes \mathbb{I}_{2} \otimes \mathbb{I}_{3} \otimes \mathbb{I}_{4}$.

Now, using the cyclic property of trace, we can write the following for an arbitrary term in Eq. (\ref{append_eq_3}),
\begin{equation}
\begin{split}
    m^{ij}_{i} &= \mathrm{Tr}\Big(\big(S_{0i}(C_{2}^{ij} \rho_{1} C_{2}^{ij^{\dagger}} \otimes \vert 0\rangle\langle 0|)S_{0i}^{\dagger}\big)M_{0}^{\prime}\Big)\nonumber \\
    & = \mathrm{Tr}\bigg( (S_{0i}^{\dagger} M_{0}^{\prime} S_{0i})(C_{2}^{ij} \rho_{1} C_{2}^{ij^{\dagger}} \otimes \vert 0\rangle\langle 0|) \bigg)\nonumber \\
    & = \mathrm{Tr}\big( M_{i}^{\prime} (C_{2}^{ij} \rho_{1} C_{2}^{ij^{\dagger}} \otimes \vert 0\rangle\langle 0|)\big)\nonumber \\
    & = \mathrm{Tr}(M_{i}C_{2}^{ij} \rho_{1} C_{2}^{ij^{\dagger}}).\mathrm{Tr}(\mathbb{I}_{0}\vert 0\rangle\langle 0\vert)\nonumber \\
    & = \mathrm{Tr}(M_{i}C_{2}^{ij} \rho_{1} C_{2}^{ij^{\dagger}}),
    \end{split}
\end{equation}
where $M_{i}\; (i \neq 0)$ denotes a Pauli-Z measurement on the $i^{\mathrm{th}}$ qubit. In this way, we design a measurement outcome which is the weighted sum of measurement outcomes from all equally probable QCNNs.\vspace{0.8em}

\bibliography{main}{}

\end{document}